\documentstyle[preprint,aps]{revtex}

\begin{document}
\draft
\title{Gauge and Lorentz Covariant Quark Propagator in an Arbitrary
Gluon Field}

\author{Yong-Hong An$^a$, Hua Yang$^{a,c}, $Qing Wang$^{a,b}$}

\address{$^a$Department of Physics, Tsinghua University, Beijing 100084, China
\footnote{Mailing address}\\
$^b$Institute of Theoretical Physics, Academia Sinica, Beijing 100080, China\\
$^c$Science School, Information Engineering University, Zhengzhou 450004}
\date{Feb 7, 2003}

\maketitle
\begin{abstract}
The quark propagator in presence of an arbitrary gluon field is 
calculated gauge and Lorentz covariantly order by order in terms
of powers of gluon field
 and its derivatives. The result is independent of path connecting ends of
 propagator and leading order result coincides with the exact
propagator in the trivial case of vanishing gluon field.
\end{abstract}

\bigskip
PACS number(s): 11.15.Tk, 12.20.Ds, 12.38.Lg, 14.65.-q

\vspace{1cm}

The quark propagator $S(x,y;A)=(i\nabla\!\!\!\! /\;-m)^{-1}(x,y)$
\footnote{Our formulae are given in Minkowski space with Bjorken Drell conventions.} with $\nabla^{\mu}=\partial^{\mu}-igA^{\mu}$ for a quark
of mass $m$ in the presence of a gluon field $A^{\mu}$ plays important role in
many investigations of quantum chromodynamics. Integrate out gluon field,
we get physical quark propagator by
\begin{eqnarray}
-i\langle 0|{\bf T}\psi(x)\overline{\psi}(y)|0\rangle
=\frac{\int{\cal D}A_{\mu}~S(x,y;A)e^{iS_{\rm QCD}(A)}}
{\int{\cal D}A_{\mu}~e^{iS_{\rm QCD}(A)}}\equiv\langle
S(x,y;A)\rangle\label{physicalS}\;,
\end{eqnarray}
where $S_{\rm QCD}(A)$ is QCD effective action with quark field be integrated 
out. Propagator $S(x,y;A)$ is Lorentz
transformation covariant and under color
gauge transformation $A_{\mu}(x)\rightarrow
V(x)A_{\mu}(x)V^{\dagger}(x)-i/gV(x)[\partial^{\mu}V^{\dagger}(x)]$ transforms
as
\begin{eqnarray}
S(x,y;A)\rightarrow S'(x,y;A)=V(x)S(x,y;A)V^{\dagger}(y)\label{Strans0}\;.
\end{eqnarray}
Except the formal expression of $S(x,y;A)$,
an expansion for $S(x,y;A)$ in terms of local gluon field $A$ is expected and
plays a key role in its applications. Note bilocal transformation law
(\ref{Strans0}) prohibit
the naive expansion ${\displaystyle\sum_n}C_n(x-y)O_n[A(x)]$ for $S(x,y;A)$
with $C_n(x-y)$ be gluon field independent coefficient and $O_n[A(x)]$ local
 operator depending on gluon field $A_{\mu}(x)$, since it is impossible for
 local operator $O_n[A(x)]$ transform bilocally. We will set up a
modified expansion by multiply naive expansion a local $A_{\mu}(x)$ dependent
"face factor" ${\bf a}[x-y;A(x)]$ with bilocal transformation law
${\bf a}[x-y;A(x)]\rightarrow {\bf a}'[x-y;A(x)]
 =V(x){\bf a}[x-y;A(x)]V^{\dagger}(y)$
\begin{eqnarray}
S(x,y;A)=\bigg[{\displaystyle\sum_n}C_n(x-y)O_n[A(x)]\bigg]~{\bf a}[x-y;A(x)]
\label{Sexp}\;
\end{eqnarray}
 which will match the transformation law (\ref{Strans0}). Then
 (\ref{physicalS}) imply we can expand physical quark propagator as,
\begin{eqnarray}
-i\langle 0|{\bf T}\psi(x)\overline{\psi}(y)|0\rangle
=\langle S(x,y;A)\rangle ={\displaystyle\sum_n}C_n(x-y)
\langle O_n[A(x)]~{\bf a}[x-y,A(x)]\rangle\;,
\end{eqnarray}
which can be treated as a gauge covariant
modified operator product expansion for quark propagator.

In literature, the most simple
approximation for the expansion of $S(x,y;A)$
is based on the perturbation expansion
\begin{eqnarray}
S(x,y;A)&=&\bigg[[1+(i\partial\!\!\! /\;-m)^{-1}gA\!\!\! /\;]^{-1}
(i\partial\!\!\! /\;-m)^{-1}\bigg](x,y)\nonumber\\
&=&(i\partial\!\!\! /\;-m)^{-1}(x,y)-
\bigg[(i\partial\!\!\! /\;-m)^{-1}gA\!\!\! /\;(i\partial\!\!\! /\;-m)^{-1}
\bigg](x,y)\nonumber\\
&&+\bigg[(i\partial\!\!\! /\;-m)^{-1}gA\!\!\! /\;(i\partial\!\!\! /\;-m)^{-1}
gA\!\!\! /\;(i\partial\!\!\! /\;-m)^{-1}\bigg](x,y)+\cdots\;,
\end{eqnarray}
the expansion can be calculated up to arbitrary orders.
 The result can be directly  expressed in terms of
powers of gluon field and its differentials, but to any fixed order of
calculation, the gauge covariance is violated.
Another approximation is so called static approximation proposed by
 Brown, Weisberger \cite{static1} and Eichten, Feinberg \cite{static2}.
By neglecting the spatial part of $\nabla\!\!\!\! /\;$, it leads
 \begin{eqnarray}
 S_{\rm static}(x,y;A)&=&-\frac{i}{2}[
 \theta(x_0-y_0)(1+\gamma^0)+\theta(y_0-x_0)(1-\gamma^0)]\nonumber\\
 &&\times
 \delta(\vec{x}-\vec{y})e^{-im|x_0-y_0|}{\bf P}e^{ig\int_y^xdz^0~A_0(z)}\;,
 \end{eqnarray}
where the path in the integral is the straight line from $y$ to $x$, 
and the path ordering is $A_0(x)$ to the left, $\cdots$, $A_0(y)$ to the right. The neglect spatial term can be subsequently taken into account as perturbation
\cite{static3}. This formalism keeps the gauge covariance of the propagator,
but Lorentz covariance is lost, it even does not coincide with the exact propagator in the trivial case of vanishing gluon field.
Recently, D.Gromes reinvestigate the problem \cite{Gromes}. He,
 in terms of path ordered exponentials, write the first order
perturbation theory as non-perturbative expression which has
correct behavior under Lorentz and gauge transformation. His result is only at  the lowest order, existence of path ordered exponentials make the expression
very complex and cause the problem of path dependence. It is purpose of present
work to invent another path independent
calculation formalism which can keep the advantages
 of different formalisms mentioned above,
\begin{itemize}
\item The expansion can be calculated up to arbitrary orders.
\item The result can be directly  expressed in terms of
powers of gluon field and its differentials and it coincide with the exact
propagator in the trivial case of vanishing gluon field.
\item The result is gauge and Lorentz transformation covariant.
\item There is no path dependence of the result.
\end{itemize}

\vspace{1cm}
We start by write quark propagator as
\begin{eqnarray}
S(x,y;A)&=&\langle x|(i\nabla\!\!\!\! /\;+m)(E-\nabla^2-m^2)^{-1}|y\rangle\;,
\label{propacal}
\end{eqnarray}
where $E-\nabla^2-m^2\equiv(i\nabla\!\!\!\! /\;-m)(i\nabla\!\!\!\! /\;+m)
=(i\nabla\!\!\!\! /\;+m)(i\nabla\!\!\!\! /\;-m)$
with $E=\frac{ig}{4}[\gamma^{\mu},\gamma^{\nu}]F_{\mu\nu}$ and
$F_{\mu\nu}\equiv \frac{i}{g}[\nabla_{\mu},\nabla_{\nu}]
=\partial_{\mu}A_{\nu}-\partial_{\nu}A_{\mu}-ig[A_{\mu},A_{\nu}]$.

The next step is to calculate matrix element
 $\langle x|(i\nabla\!\!\!\! /\;+m)(E-\nabla^2-m^2)^{-1}|y\rangle$ which in
 momentum space can be written as
\begin{eqnarray}
&&\langle x|(i\nabla\!\!\!\! /\;+m)(E-\nabla^2-m^2)^{-1}|y\rangle
\nonumber\\
&&=\int\frac{d^4k}{(2\pi)^4}
e^{-ik\cdot z}(i\nabla\!\!\!\! /\;_x+k\!\!\! /\;+m)[(k+i\nabla_x)^2
+E(x)-m^2]^{-1}1\bigg|_{z=x-y}\;.\label{starta}
\end{eqnarray}
Conventionally, directly perform Taylor expansion over operator
$\nabla_x^{\mu}$ and $E(x)$ will leads the result. The gauge covariance in this
 calculation program is not obvious, since operator $\nabla_x^{\mu}$, once
 acting on the final unity $1$, gives $-igA^{\mu}(x)$ which is not gauge
 covariant quantity. Only if
 $\nabla_x^{\mu}$ and $E(x)$ are composed of commutators such as
  $[\nabla_x^{\mu},\nabla_x^{\nu}]$ and $[\nabla_x^{\mu}, E(x)]$, the gauge
  covariance is explicit realized. We need a formalism to explicitly exhibit
  this  gauge covariance.

Consider
\begin{eqnarray}
e^{i\nabla_x\cdot\frac{\partial}{\partial k}}k^{\mu}
e^{-i\nabla_x\cdot\frac{\partial}{\partial k}}
&=&k^{\mu}+i\nabla_x^{\mu}+
F[(i\nabla_x\cdot\frac{\partial}{\partial k})d(
i\nabla_x\cdot\frac{\partial}{\partial k})](i\nabla_x^{\mu})\label{Finitial}\;.
\end{eqnarray}
where we have used relation
\begin{eqnarray}
e^ABe^{-A}&=&[e^{AdA}](B)=B+[A,B]
+\frac{1}{2!}[A,[A,B]]+
+\frac{1}{3!}[A,[A,[A,B]]]+\cdots\nonumber\\
&=&B+[A,B]+F[AdA]([A,B])\;,\nonumber
\end{eqnarray}
with $F$ function and $(AdA)^n$ operation introduced in Ref. \cite{Miller}
are defined as
\begin{eqnarray}
&&F(z)\equiv \frac{e^z-1}{z}-1
={\displaystyle\sum_{n=2}^{\infty}}\frac{z^{n-1}}{n!}\nonumber\\
&&(AdA)^0(B)\equiv B
\hspace{2cm}(AdA)^m(B)=[A,[A,\cdots,[A,B]],\cdots]\hspace{1cm}
\mbox{m times}\;.\nonumber
\end{eqnarray}
(\ref{Finitial}) can be written as
\begin{eqnarray}
k^{\mu}+i\nabla_x^{\mu}&=&
e^{i\nabla_x\cdot\frac{\partial}{\partial k}}
[k^{\mu}+\tilde{F}^{\mu}(\nabla_x,\frac{\partial}{\partial k})]
e^{-i\nabla_x\cdot\frac{\partial}{\partial k}}\;,\label{gaugecovariant}
\end{eqnarray}
in which
$\tilde{F}^{\mu}(\nabla_x,\frac{\partial}{\partial k})$ depend on
$\nabla_x^{\mu}$ and $\frac{\partial}{\partial k}$,
\begin{eqnarray}
\tilde{F}^{\mu}(\nabla_x,\frac{\partial}{\partial k})
&\equiv&
-e^{-i\nabla_x\cdot\frac{\partial}{\partial k}}
F[(i\nabla_x\cdot\frac{\partial}{\partial k})d(
i\nabla_x\cdot\frac{\partial}{\partial k})](i\nabla_x^{\mu})
e^{i\nabla_x\cdot\frac{\partial}{\partial k}}\nonumber\\
&=&
\frac{1}{2}[\nabla_x^{\nu},\nabla_x^{\mu}]\frac{\partial}{\partial k^{\nu}}
-\frac{i}{3}[\nabla_x^{\lambda},
[\nabla_x^{\nu},\nabla_x^{\mu}]]\frac{\partial^2}{\partial k^{\lambda}
\partial k^{\nu}}+O(p^4)\;.\label{Ftildeexp}
\end{eqnarray}
We find all terms in (\ref{Ftildeexp}) are gauge covariant. For
convenient of expansion we assign the each $\nabla_x^{\mu}$ a
momentum order $p$, then the terms $O(p^4)$ in (\ref{Ftildeexp})
are those commutators at least with four $\nabla^{\mu}_x$ derivatives.

In terms of $\tilde{F}$ function, (\ref{gaugecovariant}) tells us
term $k^{\mu}+i\nabla_x^{\mu}$ can be expressed in terms of 
$\frac{\partial}{\partial k}$ dependent but gauge
covariant quantity $\tilde{F}$ multiply some 
$\frac{\partial}{\partial k}$ dependent 
exponential "face factors". Apply
this to (\ref{starta}),
\begin{eqnarray}
&&\langle x|(i\nabla\!\!\!\! /\;+m)(E-\nabla^2-m^2)^{-1}|y\rangle
\nonumber\\
&&=\int\frac{d^4k}{(2\pi)^4}
e^{-ik\cdot z}
e^{i\nabla_x\cdot\frac{\partial}{\partial k}}
(\tilde{F}\!\!\!\! /\;_{\partial k}+k\!\!\! /\;+m)
e^{-i\nabla_x\cdot\frac{\partial}{\partial k}}
[e^{i\nabla_x\cdot\frac{\partial}{\partial k}}
(k+\tilde{F}_{\partial k})^2e^{-i\nabla_x\cdot\frac{\partial}{\partial k}}
+E(x)-m^2]^{-1}1\bigg|_{z=x-y}\nonumber\\
&&=\int\frac{d^4k}{(2\pi)^4}
e^{i\nabla_x\cdot\frac{\partial}{\partial k}}
(\tilde{F}\!\!\!\! /\;_{\partial k}+k\!\!\! /\;+m)
[(k+\tilde{F}_{\partial k})^2+\tilde{E}_{\partial k}(x)-m^2]^{-1}
e^{-i\nabla_x\cdot\frac{\partial}{\partial k}}
e^{-ik\cdot z}\bigg|_{z=x-y}\nonumber\\
&&=\int\frac{d^4k}{(2\pi)^4}
(\tilde{F}\!\!\!\! /\;_{\partial k}+k\!\!\! /\;+m)
[(k+\tilde{F}_{\partial k})^2+\tilde{E}_{\partial k}(x)-m^2]^{-1}e^{-ik\cdot z}
e^{-z\cdot\nabla_x}\bigg|_{z=x-y}\;,\label{mid1}
\end{eqnarray}
where $\nabla_x$ commute with $z$. In last equality, we have
dropped total momentum space derivative terms
\begin{eqnarray}
\int\frac{d^4k}{(2\pi)^4}
[e^{i\nabla_x\cdot\frac{\partial}{\partial k}}-1]\bigg[
(\tilde{F}\!\!\!\! /\;_{\partial k}+k\!\!\! /\;+m)
[(k+\tilde{F}_{\partial k})^2+\tilde{E}_{\partial k}(x)-m^2]^{-1}e^{-ik\cdot z}
e^{-z\cdot\nabla_x}\bigg]_{z=x-y}\nonumber
\end{eqnarray}
$\tilde{F}^{\mu}_{\partial k}\equiv
\tilde{F}^{\mu}(\nabla_x,\frac{\partial}{\partial k})$ and $\tilde{E}$ are all
gauge transformation covariant quantities with $\frac{\partial}{\partial k}$
 dependent $\tilde{E}$ be defined as,
\begin{eqnarray}
\tilde{E}_{\partial k}&\equiv&
\tilde{E}(\nabla_x,\frac{\partial}{\partial k})=
e^{-i\nabla_x\cdot\frac{\partial}{\partial k}}E(x)
e^{i\nabla_x\cdot\frac{\partial}{\partial k}}
\nonumber\\
&=&E(x)-i[\nabla_x^{\mu},E(x)]\frac{\partial}{\partial k^{\mu}}
-\frac{1}{2}[\nabla_x^{\nu},[\nabla_x^{\mu},E(x)]]
\frac{\partial^2}{\partial k^{\nu}\partial k^{\mu}}+\cdots\;.\label{Etildeexp}
\end{eqnarray}
(\ref{mid1}) imply
\begin{eqnarray}
S(x,y;A)&=&\tilde{S}[x-y;A(x)]~{\bf a}[x-y;A(x)]\;,\label{Sresult}
\end{eqnarray}
with  $\tilde{S}[x-y;A(x)]$ be defined as
\begin{eqnarray}
&&\tilde{S}[z;A(x)]\equiv \int\frac{d^4k}{(2\pi)^4}
[\tilde{F}\!\!\!\! /\;_{\partial k}(x)+k\!\!\! /\;+m]
[(k+\tilde{F}_{\partial k}(x))^2
+\tilde{E}_{\partial k}(x)-m^2]^{-1}e^{-ik\cdot z}
\label{tildeSdef}\\
&&{\bf a}[z;A(x)]\equiv e^{-z\cdot\nabla_x}1\;.
\end{eqnarray}
So the quark propagator in presence of gluon field is consists of
two parts: one is $\tilde{S}[x-y;A(x)]$ which can be seen as a
generalized Fourior transformation of momentum space quark
propagator in presence of local gluon field, another is a bilocal
exponential "face factor" ${\bf a}[x-y;A(x)]$. We now discuss them separately
in detail.

For $\tilde{S}[x-y;A(x)]$, note  under  gauge
transformation $V(x)$, $\nabla^{\mu}_x$ transform as
$\nabla_x^{\mu}\rightarrow V(x)\nabla_x^{\mu}V^{\dagger}(x)$ which
leads $\tilde{F}^{\mu}(\nabla_x,
\frac{\partial}{\partial k})\rightarrow
V(x)\tilde{F}^{\mu}(\nabla_x,\frac{\partial}{\partial k})V^{\dagger}(x)$ and
$\tilde{E}(\nabla_x,\frac{\partial}{\partial k})\rightarrow
V(x)\tilde{E}(\nabla_x,\frac{\partial}{\partial
k})V^{\dagger}(x)$. (\ref{tildeSdef}) then tells us
$\tilde{S}[z;A(x)]$ take the transformation rule,
\begin{eqnarray}
\tilde{S}[z;A(x)]\rightarrow V(x)\tilde{S}[z;A(x)]V^{\dagger}(x)\;.
\label{tildeStrans}
\end{eqnarray}
To calculate  $\tilde{S}[x-y;A(x)]$,
we can first expand its integrand in terms of
powers of commutators of
 $\nabla^{\mu}_x$, and denote $I_n(k,\frac{\partial}{\partial k};A)$
 be n-th order of it, i.e.
\begin{eqnarray}
&&(\tilde{F}\!\!\!\! /\;_{\partial k}+k\!\!\! /\;+m)
[(k+\tilde{F}_{\partial k})^2+\tilde{E}_{\partial k}(x)-m^2]^{-1}
={\displaystyle\sum_{n=0}}I_n(k,\frac{\partial}{\partial k};A)\;,
\end{eqnarray}
with convention that $\frac{\partial}{\partial k_{\nu}}$ is always at r.h.s. of
$k_{\nu}$. It is easy to find
\begin{eqnarray}
I_0(k,\frac{\partial}{\partial k};A)
&=&\frac{k\!\!\! /\;+m}{k^2-m^2}=\frac{1}{k\!\!\! /\;-m}\;,
\end{eqnarray}
which is just free quark propagator in momentum space. Further with help of
(\ref{Ftildeexp}) and (\ref{Etildeexp}), we find 
$I_1(k,\frac{\partial}{\partial k};A)=0$ and
\begin{eqnarray}
&&I_2(k,\frac{\partial}{\partial k};A)\nonumber\\
&&=\{\frac{i}{2}k_{\sigma}\gamma_{\rho}\gamma_5\epsilon^{\rho\sigma\mu\nu}
-\frac{m}{4}[\gamma^{\mu},\gamma^{\nu}]\}
\frac{igF_{\mu\nu}}{(k^2-m^2)^2}
+\frac{\frac{ig}{2}\gamma^{\mu}F_{\mu\nu}}{k^2-m^2}
\frac{\partial}{\partial k_{\nu}} -ig\frac{(k\!\!\! /\;+m)
F_{\mu\nu}k^{\mu}}{(k^2-m^2)^2} \frac{\partial}{\partial k_{\nu}}\\
&&I_3(k,\frac{\partial}{\partial k};A)\nonumber\\
&&=\frac{g}{3}\gamma^{\mu}[\nabla_x^{\lambda},F_{\mu\nu}]
\{\frac{-2g^{{\lambda}{\nu}}}
{(k^2-m^2)^2}+\frac{8k_{\nu}k_{\lambda}}{(k^2-m^2)^3}-\frac{2k_{\nu}}{(k^2-m^2)^2}\frac{\partial}
{\partial
k^{\lambda}}-\frac{2k_{\lambda}}{(k^2-m^2)^2}\frac{\partial}
{\partial k^{\nu}}\nonumber\\
&&\hspace{0.5cm}+\frac{1}{k^2-m^2}\frac{\partial^2}{\partial
k^{\lambda}k^{\nu}}\}
-\frac{g}{3}(k\!\!\!/\;+m)[\nabla_x^{\lambda},F_{\mu\nu}]\{
\frac{1}{(k^2-m^2)^3}[
-2g^{{\lambda}{\mu}}k^{\nu}-4g^{{\lambda}{\nu}}k^{\mu}
-4k^{\mu}k^{\lambda}\frac{\partial} {\partial k^{\nu}}]\nonumber\\
&&\hspace{0.5cm}+\frac{1}{(k^2-m^2)^2}[
g^{{\lambda}{\mu}}\frac{\partial} {\partial
k^{\nu}}+2k^{\mu}\frac{\partial^2}{\partial
k^{\lambda}k^{\nu}}]\}
+\frac{g}{2}\frac{(k\!\!\!/\;+m)k_{\lambda}}{(k^2-m^2)^{3}}
[\nabla_x^{\lambda},[\gamma^{\mu},\gamma^{\nu}]F_{\mu\nu}]
\nonumber\\
&&\hspace{0.5cm}
-\frac{g}{4}\frac{(k\!\!\!/\;+m)}{(k^2-m^2)^2}[\nabla_x^{\lambda},
[\gamma^{\mu},\gamma^{\nu}]F_{\mu\nu}]\frac{\partial}{\partial {k_{\lambda}}}\\
&&I_4(k,\frac{\partial}{\partial k};A)\nonumber\\
&&=-\frac{ig}{8}[\nabla_x^{\rho},[\nabla_x^{\lambda},F_{\mu\nu}]]
\bigg\{\frac{\gamma^\mu}{k^2-m^2}\frac{\partial^3}{\partial
k^{\rho}\partial k^{\lambda}\partial
k^{\nu}}+\frac{1}{(k^2-m^2)^{2}}\nonumber\\
&&\hspace{0.5cm}\bigg(-2\gamma^\mu(g^{\lambda\nu}
\frac{\partial}{\partial k^{\rho}}+g^{\nu\rho}\frac{\partial}{\partial
k^{\lambda}}+g^{\rho\lambda}\frac{\partial}{\partial k^{\nu}}
+k^{\rho}\frac{\partial^2}{\partial k^{\lambda}\partial k^{\nu}}
+k^{\nu}\frac{\partial^2}{\partial k^{\rho}\partial k^{\lambda}}
+k^{\lambda}\frac{\partial^2}{\partial k^{\rho}\partial k^{\nu}})\nonumber\\
&&\hspace{0.5cm}-(k\!\!\!/\;+m)(2k^{\mu}
\frac{\partial^3}{\partial k^{\rho}\partial k^{\lambda}\partial k^{\nu}}
+g^{\lambda\mu}\frac{\partial^2}{\partial k^{\rho}\partial k^{\nu}}
+g^{\rho\mu}\frac{\partial^2}{\partial k^{\lambda}\partial k^{\nu}}
+[\gamma^\mu,\gamma^\nu]
\frac{\partial^2}{\partial k^{\rho}\partial k^{\lambda}})\bigg)\nonumber\\
&&\hspace{0.5cm}+\frac{1}{(k^2-m^2)^{3}}\bigg(2(k\!\!\!/\;+m)
[g^{\rho\mu}g^{\lambda\nu}
+g^{\rho\mu}k^{\nu}\frac{\partial}{\partial k^{\lambda}}
+g^{\rho\mu}k^{\lambda}\frac{\partial}{\partial k^{\nu}}
+2k^{\mu}k^{\rho}\frac{\partial^2}{\partial k^{\lambda}\partial k^{\nu}}
\nonumber\\
&&\hspace{0.5cm}+2k^{\mu}k^{\lambda}
\frac{\partial^2}{\partial k^{\rho}\partial k^{\nu}}
+[\gamma^\mu,\gamma^\nu](g^{\rho\lambda}
+k^{\lambda}\frac{\partial}{\partial k^{\rho}}
+k^{\rho}\frac{\partial}{\partial k^{\lambda}})]
+8\gamma^{\mu}(g^{\rho\nu}k^{\lambda}+g^{\rho\lambda}k^{\nu}
+g^{\nu\lambda}k^{\rho}\nonumber\\
&&\hspace{0.5cm}+k^{\nu}k^{\lambda}
\frac{\partial}{\partial k^{\rho}}
+k^{\nu}k^{\rho}\frac{\partial}{\partial k^{\lambda}}
+k^{\rho}k^{\lambda}\frac{\partial}{\partial k^{\nu}})
+2(k\!\!\!/\;+m)(2g^{\rho\nu}k^{\mu}
\frac{\partial}{\partial k^{\lambda}}
+2g^{\rho\lambda}k^{\mu}\frac{\partial}{\partial k^{\nu}}
+2g^{\nu\lambda}k^{\mu}\frac{\partial}{\partial k^{\rho}}\nonumber\\
&&\hspace{0.5cm}+g^{\lambda\mu}g^{\rho\nu}+g^{\lambda\mu}k^{\nu}
\frac{\partial}{\partial k^{\rho}}
+g^{\lambda\mu}k^{\rho}\frac{\partial}{\partial k^{\nu}})\bigg)
+\frac{1}{(k^2-m^2)^{4}}\bigg(-48\gamma^{\mu}k^{\rho}k^{\nu}k^{\lambda}
\nonumber\\
&&\hspace{0.5cm}-8(k\!\!\!/\;+m)(2g^{\rho\nu}k^{\mu}k^{\lambda}
+2g^{\lambda\nu}k^{\mu}k^{\rho}
+g^{\lambda\mu}k^{\nu}k^{\rho}
+g^{\rho\mu}k^{\nu}k^{\lambda}
+2k^{\lambda}k^{\rho}k^{\mu}
\frac{\partial}{\partial k^{\nu}}
+[\gamma^\mu,\gamma^\nu]k^{\rho}k^{\lambda})\bigg)\bigg\}\nonumber\\
&&\hspace{0.5cm} -g^2
F_{\lambda\rho}F_{\mu\nu}\bigg\{-\frac{1}{(k^2-m^2)^2}
(\frac{1}{2}\gamma^\lambda g^{\rho\mu}
\frac{\partial}{\partial k^{\nu}}+\frac{1}{2}\gamma^\lambda k^{\mu}
\frac{\partial^2}{\partial k^{\rho}\partial k^{\nu}}
+\frac{1}{8}\gamma^\lambda[\gamma^\mu,\gamma^\nu]
\frac{\partial}{\partial k^{\rho}}\nonumber\\
&&\hspace{0.5cm}+\frac{1}{4}(k\!\!\!/\;+m)g^{\lambda\mu}
\frac{\partial^2}{\partial k^{\rho}\partial k^{\nu}})
+\frac{1}{(k^2-m^2)^3}\bigg(\gamma^\lambda(g^{\rho\mu}k^{\nu}
+g^{\rho\nu}k^{\mu}+2k^{\mu}k^{\rho}\frac{\partial}{\partial k^{\nu}}
+\frac{1}{2}k^{\rho}[\gamma^\mu,\gamma^\nu])\nonumber\\
&&\hspace{0.5cm}+\frac{1}{2}g^{\lambda\mu}(k\!\!\!/\;+m)(g^{\rho\nu}+k^{\nu}
\frac{\partial}{\partial k^{\rho}}+k^{\rho}\frac{\partial}{\partial k^{\nu}})
+k^{\lambda}(k\!\!\!/\;+m)(g^{\rho\mu}\frac{\partial}{\partial k^{\nu}}
+k^{\mu}\frac{\partial^2}{\partial k^{\rho}\partial k^{\nu}})\nonumber\\
&&\hspace{0.5cm}+\frac{(k\!\!\!/\;+m)}{4}(k^{\lambda}[\gamma^\mu,\gamma^\nu]
\frac{\partial}{\partial k^{\rho}}+k^{\mu}[\gamma^\lambda,\gamma^\rho]
\frac{\partial}{\partial k^{\nu}}+\frac{1}{4}[\gamma^\lambda,\gamma^\rho]
[\gamma^\mu,\gamma^\nu])\bigg)\nonumber\\
&&\hspace{0.5cm}-\frac{2}{(k^2-m^2)^4}(k\!\!\!/\;+m)(g^{\lambda\mu}k^{\nu}
k^{\rho}+g^{\rho\mu}k^{\nu}k^{\lambda}+g^{\rho\nu}k^{\mu}k^{\lambda})\bigg\}
\;.
\end{eqnarray}
Correspondingly we can write
\begin{eqnarray}
&&\tilde{S}[z;A(x)]={\displaystyle\sum_{n=0}}\tilde{S}_n[z;A(x)]
\label{tildeSexp}\\
&&\tilde{S}_n[z;A(x)]\equiv \int\frac{d^4k}{(2\pi)^4}
I_n(k,\frac{\partial}{\partial k};A(x))e^{-ik\cdot z}
=\int\frac{d^4k}{(2\pi)^4}I_n(k,-iz;A(x))e^{-ik\cdot z}\label{In}\;.
\end{eqnarray}
After momentum integration, (\ref{tildeSexp}) and (\ref{In})
will lead the expansion
$\tilde{S}[z;A(x)]={\displaystyle\sum_n}C_n(z)O_n[A(x)]$ mentioned previously 
in (\ref{Sexp}). The leading term
 $\tilde{S}_0[z;A(x)]=C_0(z)$ with $O_0[A(x)]=1$
 is just  the exact propagator in the trivial case of vanishing gluon field,
\begin{eqnarray}
\tilde{S}_0[z;A(x)]=\int\frac{d^4k}{(2\pi)^4}
\frac{e^{-ik\cdot z}}{k\!\!\! /\;-m}\;.
\end{eqnarray}

Now, we come to discuss the  exponential "face factor", it satisfy constraints
\begin{eqnarray}
&&{\bf a}[0;A(x)]=1\hspace{2cm}
{\bf a}[x-y;0]=1\nonumber\\
&&(x-y)\cdot\nabla_x{\bf a}[x-y;A(x)]=
[z\cdot\nabla_x+z\cdot\partial_z]e^{-z\cdot\nabla_x}1\bigg|_{z=x-y}=0\;.
\end{eqnarray}
In the literature, these constraints usually lead to path ordered
non-integratable face factor ${\bf P}e^{ig\int_y^xdz^{\mu}A_{\mu}(z)}$
\cite{Ball} which depend on
path. Our result instead only rely on end points $x,y$ and is
independent of the path. Except the formal definition of
${\bf a}[x-y;A(x)]$,
 the explicit expression of ${\bf a}[x-y;A(x)]$ can be got with help of
 Baker-Housdoff formula
\begin{eqnarray}
{\bf a}[x-y;A(x)]&=&
\bigg[e^{-z\cdot\nabla_x}e^{z\cdot\partial_x}1\bigg]_{z=x-y}
\equiv e^{C(x,z)}\bigg|_{z=x-y}\;,
\end{eqnarray}
where $C(x,z)$ is defined as
\begin{eqnarray}
e^{C(x,z)}&=&e^{-z\cdot\nabla_x}e^{z\cdot\partial_x}
=\exp\bigg[-z\cdot\nabla_x+z\cdot\partial_x
+\frac{1}{2}[-z\cdot\nabla_x,z\cdot\partial_x]
+\frac{1}{12}[-z\cdot\nabla_x,[-z\cdot\nabla_x,z\cdot\partial_x]]\nonumber\\
&&-\frac{1}{12}[z\cdot\partial_x,[z\cdot\partial_x,-z\cdot\nabla_x]]
+\cdots\bigg]
\nonumber\\
&=&\exp\bigg[igz\cdot A(x)
+\frac{1}{2}[igz\cdot A(x),z\cdot\partial_x]
+\frac{1}{12}[igz\cdot\nabla_x,[iz\cdot A(x),z\cdot\partial_x]]\nonumber\\
&&-\frac{1}{12}[z\cdot\partial_x,[z\cdot\partial_x,igz\cdot A(x)]]+\cdots\bigg]
\;.\label{Cdef}
\end{eqnarray}
Note $C(x,z)$ donot include pure operator $\partial_x$ again, all $\partial_x$
in  $C(x,z)$ are already acting on gluon filed $A_{\mu}(x)$ and gluon field
dependence in $C(x,z)$ is local at space-time point $x$.

The gauge transformation covariance of ${\bf a}[x-y;A(x)]$ can be proved as
follows: since  under  gauge transformation $V(x)$, $\nabla^{\mu}_x$ transform
as $\nabla_x^{\mu}\rightarrow V(x)\nabla_x^{\mu}V^{\dagger}(x)$,
 ${\bf a}[x-y;A(x)]$ then transform as
\begin{eqnarray}
{\bf a}[x-y;A(x)]&\rightarrow&\bigg[e^{-V(x)z\cdot\nabla_xV^{\dagger}(x)}1
\bigg]_{z=x-y}
=V(x)\bigg[e^{-z\cdot\nabla_x}V^{\dagger}(x)
\bigg]_{z=x-y}\nonumber\\
&=&V(x)\bigg[e^{-z\cdot\nabla_x}
e^{z\cdot\partial_x}e^{-z\cdot\partial_x}V^{\dagger}(x)
\bigg]_{z=x-y}
=V(x)e^{C(x,z)}
\bigg[e^{-z\cdot\partial_x}V^{\dagger}(x)\bigg]_{z=x-y}\nonumber\\
&=&\bigg[V(x)e^{C(x,z)}V^{\dagger}(x-z)\bigg]_{z=x-y}
=V(x){\bf a}(x,y;A)V^{\dagger}(y)\;,\label{atrans}
\end{eqnarray}
where we have used property
\begin{eqnarray}
e^{-z\cdot\partial_x}V^{\dagger}(x)&=&
\bigg[1+{\displaystyle\sum_{n=1}^{\infty}}\frac{(-1)^n}{n!}
z_{\mu_1}\cdots z_{\mu_n}\partial_{x,\mu_1}\cdots\partial_{x,\mu_n}\bigg]
V^{\dagger}(x)=V^{\dagger}(x-z)\nonumber\;.
\end{eqnarray}
Combine (\ref{tildeStrans}) and (\ref{atrans}) together, we find our result
quark propagator (\ref{Sresult}) is gauge transformation covariant
\begin{eqnarray}
S(x,y;A)\rightarrow V(x)S(x,y;A)V^{\dagger}(y)\;.\label{Strans}
\end{eqnarray}
Similarly since the Lorentz covariance for $\tilde{S}$ and ${\bf a}$ is
explicit, our result propagator is explicit Lorentz covariant.

In conclusion, we have factorized the quark propagator $S(x,y;A)$ by a
generalized Fourior transformation of  momentum space quark propagator
$\tilde{S}[x-y;A(x)]$
in presence of gluon field and a path independent "face factor"
${\bf a}[x-y;A(x)]$. The two parts are all
only depend on gluon field at local space-time point $x$. The
formalism is gauge and Lorentz covariant, it coincide with the exact
propagator in the trivial case of vanishing gluon field.

\section*{Acknowledgments}

This work was  supported by National  Science Foundation of China No.90103008
and fundamental research grant of Tsinghua University.



\end{document}